\documentstyle[12pt]{article}
\makeatletter
\let\DOTSI\relax
\def\RIfM@{\relax\ifmmode}%
\def\FN@{\futurelet\next}%
\newcount\intno@
\def\iint{\DOTSI\intno@\tw@\FN@\ints@}%
\def\iiint{\DOTSI\intno@\thr@@\FN@\ints@}%
\def\iiiint{\DOTSI\intno@4 \FN@\ints@}%
\def\idotsint{\DOTSI\intno@\z@\FN@\ints@}%
\def\ints@{\findlimits@\ints@@}%
\newif\iflimtoken@
\newif\iflimits@
\def\findlimits@{\limtoken@true\ifx\next\limits\limits@true
 \else\ifx\next\nolimits\limits@false\else
 \limtoken@false\ifx\ilimits@\nolimits\limits@false\else
 \ifinner\limits@false\else\limits@true\fi\fi\fi\fi}%
\def\multint@{\int\ifnum\intno@=\z@\intdots@                                %1
 \else\intkern@\fi                                                          %2
 \ifnum\intno@>\tw@\int\intkern@\fi                                         %3
 \ifnum\intno@>\thr@@\int\intkern@\fi                                       %4
 \int}%                                                                     %5
\def\multintlimits@{\intop\ifnum\intno@=\z@\intdots@\else\intkern@\fi
 \ifnum\intno@>\tw@\intop\intkern@\fi
 \ifnum\intno@>\thr@@\intop\intkern@\fi\intop}%
\def\intic@{\mathchoice{\hskip.5em}{\hskip.4em}{\hskip.4em}{\hskip.4em}}%
\def\negintic@{\mathchoice
 {\hskip-.5em}{\hskip-.4em}{\hskip-.4em}{\hskip-.4em}}%
\def\ints@@{\iflimtoken@                                                    %1
 \def\ints@@@{\iflimits@\negintic@\mathop{\intic@\multintlimits@}\limits    %2
  \else\multint@\nolimits\fi                                                %3
  \eat@}%                                                                   %4
 \else                                                                      %5
 \def\ints@@@{\iflimits@\negintic@
  \mathop{\intic@\multintlimits@}\limits\else
  \multint@\nolimits\fi}\fi\ints@@@}%
\def\intkern@{\mathchoice{\!\!\!}{\!\!}{\!\!}{\!\!}}%
\def\plaincdots@{\mathinner{\cdotp\cdotp\cdotp}}%
\def\intdots@{\mathchoice{\plaincdots@}%
 {{\cdotp}\mkern1.5mu{\cdotp}\mkern1.5mu{\cdotp}}%
 {{\cdotp}\mkern1mu{\cdotp}\mkern1mu{\cdotp}}%
 {{\cdotp}\mkern1mu{\cdotp}\mkern1mu{\cdotp}}}%
\def\rmfam{\z@}%
\newif\iffirstchoice@
\firstchoice@true
\def\textfonti{\the\textfont\@ne}%
\def\textfontii{\the\textfont\tw@}%
\def\text{\RIfM@\expandafter\text@\else\expandafter\text@@\fi}%
\def\text@@#1{\leavevmode\hbox{#1}}%
\def\text@#1{\mathchoice
 {\hbox{\everymath{\displaystyle}\def\textfonti{\the\textfont\@ne}%
  \def\textfontii{\the\textfont\tw@}\textdef@@ T#1}}%
 {\hbox{\firstchoice@false
  \everymath{\textstyle}\def\textfonti{\the\textfont\@ne}%
  \def\textfontii{\the\textfont\tw@}\textdef@@ T#1}}%
 {\hbox{\firstchoice@false
  \everymath{\scriptstyle}\def\textfonti{\the\scriptfont\@ne}%
  \def\textfontii{\the\scriptfont\tw@}\textdef@@ S\rm#1}}%
 {\hbox{\firstchoice@false
  \everymath{\scriptscriptstyle}\def\textfonti
  {\the\scriptscriptfont\@ne}%
  \def\textfontii{\the\scriptscriptfont\tw@}\textdef@@ s\rm#1}}}%
\def\textdef@@#1{\textdef@#1\rm\textdef@#1\bf\textdef@#1\sl\textdef@#1\it}%
\def\DN@{\def\next@}%
\def\eat@#1{}%
\def\textdef@#1#2{%
 \DN@{\csname\expandafter\eat@\string#2fam\endcsname}%
 \if S#1\edef#2{\the\scriptfont\next@\relax}%
 \else\if s#1\edef#2{\the\scriptscriptfont\next@\relax}%
 \else\edef#2{\the\textfont\next@\relax}\fi\fi}%
\def\Let@{\relax\iffalse{\fi\let\\=\cr\iffalse}\fi}%
\def\vspace@{\def\vspace##1{\crcr\noalign{\vskip##1\relax}}}%
\def\multilimits@{\bgroup\vspace@\Let@
 \baselineskip\fontdimen10 \scriptfont\tw@
 \advance\baselineskip\fontdimen12 \scriptfont\tw@
 \lineskip\thr@@\fontdimen8 \scriptfont\thr@@
 \lineskiplimit\lineskip
 \vbox\bgroup\ialign\bgroup\hfil$\m@th\scriptstyle{##}$\hfil\crcr}%
\def\Sb{_\multilimits@}%
\def\endSb{\crcr\egroup\egroup\egroup}%
\def\Sp{^\multilimits@}%
\let\endSp\endSb
\newdimen\ex@
\def\rightarrowfill@#1{$#1\m@th\mathord-\mkern-6mu\cleaders
 \hbox{$#1\mkern-2mu\mathord-\mkern-2mu$}\hfill
 \mkern-6mu\mathord\rightarrow$}%
\def\leftarrowfill@#1{$#1\m@th\mathord\leftarrow\mkern-6mu\cleaders
 \hbox{$#1\mkern-2mu\mathord-\mkern-2mu$}\hfill\mkern-6mu\mathord-$}%
\def\leftrightarrowfill@#1{$#1\m@th\mathord\leftarrow\mkern-6mu\cleaders
 \hbox{$#1\mkern-2mu\mathord-\mkern-2mu$}\hfill
 \mkern-6mu\mathord\rightarrow$}%
\def\overrightarrow{\mathpalette\overrightarrow@}%
\def\overrightarrow@#1#2{\vbox{\ialign{##\crcr\rightarrowfill@#1\crcr
 \noalign{\kern-\ex@\nointerlineskip}$\m@th\hfil#1#2\hfil$\crcr}}}%

\def\overleftarrow{\mathpalette\overleftarrow@}%
\def\overleftarrow@#1#2{\vbox{\ialign{##\crcr\leftarrowfill@#1\crcr
 \noalign{\kern-\ex@\nointerlineskip}$\m@th\hfil#1#2\hfil$\crcr}}}%
\def\overleftrightarrow{\mathpalette\overleftrightarrow@}%
\def\overleftrightarrow@#1#2{\vbox{\ialign{##\crcr\leftrightarrowfill@#1\crcr
 \noalign{\kern-\ex@\nointerlineskip}$\m@th\hfil#1#2\hfil$\crcr}}}%
\def\underrightarrow{\mathpalette\underrightarrow@}%
\def\underrightarrow@#1#2{\vtop{\ialign{##\crcr$\m@th\hfil#1#2\hfil$\crcr
 \noalign{\nointerlineskip}\rightarrowfill@#1\crcr}}}%

\def\underleftarrow{\mathpalette\underleftarrow@}%
\def\underleftarrow@#1#2{\vtop{\ialign{##\crcr$\m@th\hfil#1#2\hfil$\crcr
 \noalign{\nointerlineskip}\leftarrowfill@#1\crcr}}}%
\def\underleftrightarrow{\mathpalette\underleftrightarrow@}%
\def\underleftrightarrow@#1#2{\vtop{\ialign{##\crcr$\m@th\hfil#1#2\hfil$\crcr
 \noalign{\nointerlineskip}\leftrightarrowfill@#1\crcr}}}%
\def\stackunder#1#2{\mathrel{\mathop{#2}\limits_{#1}}}%
\newcount\GRAPHICSTYPE
\GRAPHICSTYPE=\z@
\def\GRAPHICSPS#1{%
 \ifcase\GRAPHICSTYPE%\GRAPHICSTYPE=0
  ps: #1%
 \or%\GRAPHICSTYPE=1
  language "PS", include "#1"%
 \or%\GRAPHICSTYPE=2
  #1%
 \fi
}%
\def\graffile#1#2#3#4{%
 \ifnum\GRAPHICSTYPE=\tw@
  \@ifundefined{psfig}{\input psfig.tex}{}%
  \psfig{file=#1, height=#3, width=#2}%
 \else
  \leavevmode\raise -#4 \hbox{%
   \raise #3 \hbox{\rule{0.003in}{0.003in}\special{#1}}%
   }%
  {\raise -#4 \hbox to #2 {\vrule height#3 width\z@ depth\z@\hfil}}%
 \fi
}%
\def\draftbox#1#2#3#4{%
 \leavevmode\raise -#4 \hbox{%
  \frame{\rlap{\protect\tiny #1}\hbox to #2%
   {\vrule height#3 width\z@ depth\z@\hfil}%
  }%
 }%
}%
\newcount\draft
\draft=\z@
\def\GRAPHIC#1#2#3#4#5{%
 \ifnum\draft=\@ne\draftbox{#2}{#3}{#4}{#5}%
  \else\graffile{#1}{#3}{#4}{#5}%
  \fi
 }%
\def\addtoLaTeXparams#1{\edef\LaTeXparams{\LaTeXparams #1}}%
\def\doFRAMEparams#1{\readFRAMEparams#1\end}%
\def\readFRAMEparams#1{%
 \ifx#1\end%
  \let\next=\relax
  \else
  \ifx#1i\dispkind=\z@\fi
  \ifx#1d\dispkind=\@ne\fi
  \ifx#1f\dispkind=\tw@\fi
  \ifx#1t\addtoLaTeXparams{t}\fi
  \ifx#1b\addtoLaTeXparams{b}\fi
  \ifx#1p\addtoLaTeXparams{p}\fi
  \ifx#1h\addtoLaTeXparams{h}\fi
  \let\next=\readFRAMEparams
  \fi
 \next
 }%
\def\IFRAME#1#2#3#4#5{\GRAPHIC{#5}{#4}{#1}{#2}{#3}}%
\def\DFRAME#1#2#3#4{%
 \begin{center}\GRAPHIC{#4}{#3}{#1}{#2}{\z@}\end{center}%
 }%
\def\FFRAME#1#2#3#4#5#6#7{%
 \begin{figure}[#1]%
  \begin{center}\GRAPHIC{#7}{#6}{#2}{#3}{\z@}\end{center}%
  \caption{\label{#5}#4}%
  \end{figure}%
 }%
\newcount\dispkind%
\def\FRAME#1#2#3#4#5#6#7#8{%
 \def\LaTeXparams{}%
 \dispkind=\z@
 \def\LaTeXparams{}%
 \doFRAMEparams{#1}%
 \ifnum\dispkind=\z@\IFRAME{#2}{#3}{#4}{#7}{#8}\else
  \ifnum\dispkind=\@ne\DFRAME{#2}{#3}{#7}{#8}\else
   \ifnum\dispkind=\tw@
    \edef\@tempa{\noexpand\FFRAME{\LaTeXparams}}%
    \@tempa{#2}{#3}{#5}{#6}{#7}{#8}%
    \fi
   \fi
  \fi
 }%
\long\def\QQQ#1#2{\long\expandafter\def\csname#1\endcsname{#2}}%
\def\QTP#1{}%
\long\def\QQA#1#2{}%
\def\QTR#1#2{{\csname#1\endcsname #2}}%(gp) Is this the best?
\def\EXPAND#1[#2]#3{}%
\def\NOEXPAND#1[#2]#3{}%
\def\LaTeXparent#1{}%
\def\ChildStyles#1{}%
\def\ChildDefaults#1{}%
\def\QTagDef#1#2#3{}%
\def\QQfnmark#1{\footnotemark}

\@ifundefined{INDEX}{\def\INDEX#1#2{}{}}{}%
\@ifundefined{SUBINDEX}{\def\SUBINDEX#1#2#3{}{}{}}{}%
\def\initial#1{\bigbreak{\raggedright\large\bf #1}\kern 2\p@\penalty3000}%
\@ifundefined{abstract}{%
 \def\abstract{%
  \if@twocolumn
   \section*{Abstract (Not appropriate in this style!)}%
   \else \small 
   \begin{center}{\bf Abstract\vspace{-.5em}\vspace{\z@}}\end{center}%
   \quotation 
   \fi
  }%
 }{%
 }%
\@ifundefined{endabstract}{\def\endabstract
  {\if@twocolumn\else\endquotation\fi}}{}%
\@ifundefined{maketitle}{\def\maketitle#1{}}{}%
\@ifundefined{affiliation}{\def\affiliation#1{}}{}%
\@ifundefined{proof}{}{}%
\@ifundefined{endproof}{}{}%
\@ifundefined{newfield}{\def\newfield#1#2{}}{}%
\@ifundefined{chapter}{\def\chapter#1{\par(Chapter head:)#1\par }%
 \newcount\c@chapter}{}%
\@ifundefined{part}{\def\part#1{\par(Part head:)#1\par }}{}%
\@ifundefined{section}{\def\section#1{\par(Section head:)#1\par }}{}%
\@ifundefined{subsection}{\def\subsection#1%
 {\par(Subsection head:)#1\par }}{}%
\@ifundefined{subsubsection}{\def\subsubsection#1%
 {\par(Subsubsection head:)#1\par }}{}%
\@ifundefined{paragraph}{\def\paragraph#1%
 {\par(Subsubsubsection head:)#1\par }}{}%
\@ifundefined{subparagraph}{\def\subparagraph#1%
 {\par(Subsubsubsubsection head:)#1\par }}{}%
\@ifundefined{therefore}{}{}%
\@ifundefined{backepsilon}{}{}%
\@ifundefined{yen}{}{}%
\@ifundefined{registered}{\def\registered{\relax\ifmmode{}\r@gistered
                                                \else$\m@th\r@gistered$\fi}%
 \def\r@gistered{^{\ooalign
  {\hfil\raise.07ex\hbox{$\scriptstyle\rm\text{R}$}\hfil\crcr
  \mathhexbox20D}}}}{}%
\@ifundefined{Eth}{}{}%
\@ifundefined{eth}{}{}%
\@ifundefined{Thorn}{}{}%
\@ifundefined{thorn}{}{}%
\@ifundefined{degree}{}{}%
\def\BibTeX{{\rm B\kern-.05em{\sc i\kern-.025em b}\kern-.08em
    T\kern-.1667em\lower.7ex\hbox{E}\kern-.125emX}}%
\newdimen\theight
\def\Column{%
 \vadjust{\setbox\z@=\hbox{\scriptsize\quad\quad tcol}%
  \theight=\ht\z@\advance\theight by \dp\z@\advance\theight by \lineskip
  \kern -\theight \vbox to \theight{%
   \rightline{\rlap{\box\z@}}%
   \vss
   }%
  }%
 }%
\def\qed{%
 \ifhmode\unskip\nobreak\fi\ifmmode\ifinner\else\hskip5\p@\fi\fi
 \hbox{\hskip5\p@\vrule width4\p@ height6\p@ depth1.5\p@\hskip\p@}%
 }%
\def\miss{\hbox{\vrule height2\p@ width 2\p@ depth\z@}}%
\def\tcol#1{{\baselineskip=6\p@ \vcenter{#1}} \Column}  %
\makeatother
\begin{document}

\title{{\small \noindent hep-th/9604200\hfill USC-96/HEP-B3}\bigskip\\
Duality and hidden dimensions\thanks{%
Lecture delivered in honor of Keiji Kikkawa for his 60th birtday at the
conference Frontiers in Quantum Field Theory, Toyonaka, Osaka, Japan,
Dec.1996.}}
\author{Itzhak Bars\thanks{%
Research supported by DOE grant No. DE-FG03-44ER-40168} \\
%EndAName
Department of Physics and Astronomy\\
University of Southern California\\
Los Angeles, CA 90089-0484, USA\bigskip}
\date{April 30, 1996}
\maketitle

\begin{abstract}
Using a global superalgebra with 32 fermionic and 528 bosonic charges, many
features of p-brane dualities and hidden dimensions are discussed.
\end{abstract}

\section{Introduction}

It is a pleasure for me to participate in the celebration of Keiji Kikkawa's
birthday. Considering that Kikkawa and Yamasaki \cite{kikkawayam} were the
first to notice T-duality, and the first to discuss membranes in the context
of unification \cite{kikkawamem}, it is only appropriate that I concentrate
my discussion on duality in strings, membranes and more generally $p$-branes.

The discovery of string dualities \cite{dualities} 
 \thinspace 
\cite{witten} have led to
the idea that there is a more fundamental theory than string theory,
``M-theory'' \cite{townsend11} \thinspace 
\cite{jhs} \thinspace 
\cite{horavawitten}, 
that manifests itself in
different forms in certain regimes of its moduli space. The several familiar
string theories (type-I, type-II, heterotic) may be regarded as different
starting points for perturbative expansions around some vacuua of the
fundamental theory, in analogy to perturbative expansions around the
different vacuua of spontaneously broken gauge theories. A lot of evidence
has accumulated by now to convince oneself that the different versions of $%
D=10$ superstrings and their compactifications are related to each other
non-perturbatively by duality transformations. Furthermore, there is
evidence that the non-perturbative theory is hiding higher dimensions and
that it is related to various $p$-branes  \cite{townsend} 
 and $D$-branes \cite{pol}.

Although I refer to ``M-theory'' I will not discuss it directly. Instead,
without going into the details of string theory or M-theory, I will connect
the duality and 11D (or even 12D) properties to a superalgebra involving 32
supercharges and 528 bosonic generators \cite{townsend} 
. Therefore, I will begin my
discussion by outlining some of the properties of the superalgebra and the
interpretation of its structure. I will then discuss examples of how
U-duality and hidden spacetime dimensions become manifest in the
non-perturbative spectrum of the theory. For more details see \cite
{ibbeyondm} for the first part and \cite{ib11d} \thinspace
\cite{ibyank} for the second
part.

\section{Dynamical superalgebra and p-branes}

It is well known that the maximum number of supercharges in a physical
theory is 32. This constraint is obtained in four dimensions by requiring
that supermultiplets of massless particles should not contain spins that
exceed 2. Assuming that the four dimensional theory is related to a higher
dimensional one, then the higher theory can have at most 32 real
supercharges. The supersymmetry associated with these supercharges is not
necessarily exact; it may be broken by central extensions included in the
superalgebra. Denote the 32 supercharges by $Q_\alpha ^a$, where $%
a=1,2,\cdots N$, and $\alpha $ is the spinor index in $d$-dimensions. For
example, in $d=11$ there is a single 32-component Majorana spinor (N=1), in
D=10 there are two 16-component Majorana-Weyl spinors (N=2), etc. down to
D=4 where there are eight 4-component Majorana spinors (N=8). It is
important to note that 32 corresponds to counting {\it real} components of
spinors.

In 12 dimensions the Weyl spinor also has 32 components since $\frac
122^{12/2}=32,$ but when the signature is $\left( 11,1\right) $ the spinor
is complex and has 64 real components. Therefore, as long as we consider a
single time coordinate, $d=11$ is the highest allowed dimension. However, if
the signature is $\left( 10,2\right) ,$ it is possible to impose a Majorana
condition that permits a real 32-component spinor. Beyond 12 dimensions the
spinor is too large, and therefore we cannot consider $d>12$.

The 32 spinors $Q_\alpha ^a$ may be classified as the spinor for $%
SO(c+1,1)\otimes SO(d-1,1)$ with $d+c+2=12.$ Here $c$ is interpreted as the
number of compactified dimensions from the point of view of 10D string
theory, and the extra 2 dimensions are considered hidden. This spinor$\times 
$ spinor classification is given in Table I for each dimension. The index $a$
corresponds to the spinor of $SO(c+1,1).$ This group is not necessarily a
symmetry, but it helps to keep track of the compactified dimensions,
including the hidden ones. Furthermore, the same index $a$ will be
reclassified later under the maximal compact subgroup $K$ of $U$-duality,
thus providing a bridge between duality and higher hidden dimensions.

Consider the maximally extended algebra of the 32 supercharges in various
dimensions in the form 
\begin{equation}
\left\{ Q_\alpha ^a,Q_\beta ^b\right\} =\delta ^{ab}\gamma _{\alpha \beta
}^\mu \,\,P_\mu +\sum_{p=0,1,\cdots }\gamma _{\alpha \beta }^{\mu _1\cdots
\mu _p}\,\,\,Z_{\mu _1\cdots \mu _p}^{ab}.  \label{supergeneral}
\end{equation}
Since the left side is the symmetric product of 32 supercharges, the right
side can have at most $\frac 1232\times 33=528$ independent generators. The
indices $ab$ on $Z_{\mu _1\cdots \mu _p}^{ab}$ are either symmetrized or
antisymmetrized and have the same permutation symmetry as $\alpha \beta $ in 
$\gamma _{\alpha \beta }^{\mu _1\cdots \mu _p}.$ The central extensions $%
Z_{\mu _1\cdots \mu _p}^{ab}\,$ are assumed to commute with $Q_\alpha
^a,P_\mu ,$ but they are tensors of the Lorentz group and hence do not
commute with it.

In (10,2) dimensions we will use $M=0^{\prime },0,1,2,\cdots ,10$ for the
space index instead of $\mu .$ In the 32$\times 32$ representation
(equivalent to chirally projected 64$\times 64$) only the 2- and 6- index
gamma matrices $\gamma _{\alpha \beta }^{M_1M_2}$ and $\gamma _{\alpha \beta
}^{M_1\cdots M_6}$ are symmetric in $\alpha \beta ,$ and furthermore $\gamma
_{\alpha \beta }^{M_1\cdots M_6}$ is self dual. Therefore, in 12 dimensions,
on the right hand side of (\ref{supergeneral}) there can be no $P_M,$ and
the 528 generators consist of $Z_{M_1M_2}$, and the self dual $Z_{M_1\cdots
M_6}^{+}.$ The number of components in each is $\frac{12\times 11}2=66$ and $%
\frac 12\frac{12\times 11\times 10\times 9\times 8\times 7}{1\times 2\times
3\times 4\times 5\times 6}=462$ respectively. Upon compactification to
(10,1) we rewrite the 12D index $M=(0^{\prime },\mu )$ where $\mu
=0,1,2,\cdots ,10$ is an 11D index. Then we have (suppressing the $0^{\prime
}$ index) 
\begin{eqnarray}
Z_{M_1M_2} &\rightarrow &P_\mu \oplus Z_{\mu _1\mu _2}\quad 66=11+55 \\
Z_{M_1\cdots M_6}^{+} &\rightarrow &X_{\mu _1\cdots \mu _5}\quad \quad \quad
462=462  \nonumber
\end{eqnarray}
which are the momenta and central charges in 11 dimensions pointed out in 
\cite{townsend}.

Continuing the compactification process to lower dimensions on $%
R^{d-1,1}\otimes T^{c+1,1},$ each eleven dimensional index $\mu $ decomposes
into $\mu \oplus m$ where $\mu $ is in $d$ dimensions and $m$ is in $%
c+1=11-d $ dimensions. 
Then each 11 dimensional tensor decomposes as follows 
\begin{eqnarray}
P_\mu &\rightarrow &P_\mu \oplus P_m  \nonumber \\
Z_{\mu \nu } &\rightarrow &Z_{\mu \nu }\oplus Z_\mu ^n\oplus Z^{mn} 
\nonumber \\
X_{\mu _1\cdots \mu _5} &\rightarrow &X_{\mu _1\cdots \mu _5}\oplus X_{\mu
_1\cdots \mu _4}^{m_1}\oplus X_{\mu _1\mu _2\mu _3}^{m_1m_2} \\
&&\oplus X_{\mu _1\mu _2}^{m_1m_2m_3}\oplus X_{\mu _1}^{m_1\cdots m_4}\oplus
X^{m_1\cdots m_5}.  \nonumber
\end{eqnarray}
For example in $d=10$ the type IIA superalgebra is recovered, with the 528
operators ($P_\mu ,P_{10},Z_{\mu \nu },Z_\mu ,X_{\mu _1\cdots \mu _4},X_{\mu
_1\cdots \mu _5}^{\pm })$ where the $\pm $ indicate self/antiself dual
respectively. In Table I in each row labelled by $(d-1,1)/(c+1,1)$ the
numbers of each central extension of $P,Z,X$ type with $p$ Lorentz indices
is indicated (these are the numbers that are not in bold type). As we go to
lower dimensions one must use the duality between $p$ indices and $d-p$
indices to reclassify and count the central extensions $Z_{\mu _1\cdots \mu
_p}^{ab}$. In the table a number in parenthesis means that it should be
omitted from there and instead moved in the same row to the location where
the same number appears in brackets . This corresponds to the equivalence of 
$p$ indices and $d-p$ indices. When $p=d-p$ there are self-dual or
anti-self-dual tensors. Their numbers are indicated with additional
superscripts $\pm $ in the form $1^{\pm },2^{\pm },3^{\pm },10^{\pm
},35^{\pm }$ wherever they occur.

The total number of central extensions $P,Z,X$ found according to this
compactification procedure for each value of $p$ are indicated in Table-I in
bold characters. These totals are the same numbers found by counting the 
{\it number} of possibilities $ab$ on $Z_{\mu _1\cdots \mu _p}^{ab}.$ The
bold numbers following the $=$ sign correspond to representations of $%
SO(c+1,1)$ (making a connection to 12D) and those following the $\approx $
sign correspond to representations of $K$ (to be discussed later in
connection to duality).

\[
\begin{tabular}{|c|c|c|c|c|c|c|c|c|}
\hline
$^{\frac{c+1,1}{d-1,1}}$ & $_{\Sp SO(c+1,1)  \\ (\text{or\thinspace }%
K\,)\,\,\otimes \,\,  \\ SO(d-1,1)  \endSp }^{32\,\,Q_\alpha ^a}$ & $\Sb %
P^m,Z^{mn}  \\ X^{mnlqr}  \endSb \Sp p=0  \\ .  \endSp $ & $\Sb P_\mu ,Z_\mu
^n  \\ X_\mu ^{nlqr}  \endSb \Sp p=1  \\ .  \endSp $ & $\Sb Z_{\mu v}  \\ %
X_{\mu \nu }^{lqr}  \endSb \Sp p=2  \\ .  \endSp $ & $\Sb .  \\ X_{\mu \nu
\lambda }^{qr}  \endSb \Sp p=3  \\ .  \endSp $ & $\Sb .  \\ X_{_{\mu _1.\mu
_4}}^r  \endSb \Sp p=4  \\ .  \endSp $ & $\Sb .  \\ X_{\mu _1.\mu _5} 
\endSb \Sp p=5  \\ .  \endSp $ & $\frac{^U}K$ \\ \hline
$\Sp A  \\ \frac{1,1}{9,1}  \endSp $ & $^{\left( {\bf \pm ,16}\right) }$ & $
\Sp 1+0  \\ +0  \endSp $ & $\Sp 1+1  \\ +0  \endSp $ & $\Sp 1  \\ +0  \endSp 
$ & $^0$ & $^1$ & $\Sp 1^{+}  \\ +1^{-}  \endSp $ & $\frac{^{SO(1,1)}}{Z_2}$
\\ \hline
$\Sp B  \\ \frac{1,1}{9,1}  \endSp $ & $^{\left( _{+}^{+}{\bf ,16}\right) }$
& $\Sp 0+0  \\ +0  \endSp $ & $\Sp 1+2  \\ +0  \endSp $ & $\Sp 0  \\ +0 
\endSp $ & $^1$ & $^0$ & $\Sp 1^{+}  \\ +2^{+}  \endSp $ & $\frac{^{SL(2,R)}%
}{^{SO(2)}}$ \\ \hline
$^{\frac{2,1}{8,1}}$ & $^{\left( {\bf 2,16}\right) }$ & $\Sp 2+1  \\ +0  \\ =%
{\bf 3}  \\ \approx {\bf 2+1}  \endSp $ & $\Sp 1+2  \\ +0  \\ ={\bf 3}  \\ %
\approx {\bf 2+1}  \endSp $ & $\Sp 1+0  \\ ={\bf 1}  \\ \approx {\bf 1} 
\endSp $ & $^1$ & $\Sp \left[ 1\right]  \\ +2  \\ ={\bf 3}  \\ \approx {\bf 2%
}  \\ {\bf +1}  \endSp $ & $\Sp \left( 1\right)  \\ _{move}  \endSp $ & $%
\frac{\Sp SL(2)\otimes  \\ SO(1,1)  \endSp }{\Sp SO(2)  \\ \otimes Z_2 
\endSp }$ \\ \hline
$^{\frac{3,1}{7,1}}$ & $^{\Sp \left( \left( {\bf 2},0\right) {\bf ,8}%
^{+}\right)  \\ \left( \left( 0,{\bf 2}\right) {\bf ,8}^{-}\right)  \endSp }$
& $\Sp 3+3  \\ +0  \\ ={\bf 6}  \\ \approx {\bf 3}^{{\bf +}}  \\ +{\bf 3}%
^{-}  \endSp $ & $\Sp 1+3  \\ +0  \\ =\left( {\bf 2,2}\right)  \\ \approx 
{\bf 3+1}  \endSp $ & $\Sp 1+1  \\ ={\bf 1+1}  \\ \approx {\bf 1+1}  \endSp $
& $\Sp 3+\left[ 1\right]  \\ =\left( {\bf 2,2}\right)  \\ \approx {\bf 3+1} 
\endSp $ & $\Sp 3^{+}  \\ +3^{-}  \\ ={\bf 6}  \\ \approx {\bf 3}^{+}  \\ +%
{\bf 3}^{-}  \endSp $ & $\Sp \left( 1\right)  \\ _{move}  \endSp $ & $^{%
\frac{\Sp SL(3)  \\ \otimes SL(2)  \endSp }{\Sp SO(3)  \\ \otimes U(1) 
\endSp }}$ \\ \hline
$^{\frac{4,1}{6,1}}$ & $^{\left( {\bf 4,8}\right) }$ & $\Sp 4+6  \\ +0  \\ =%
{\bf 10}  \\ \approx {\bf 10}  \endSp $ & $\Sp 1+4  \\ +1  \\ ={\bf 5+1}  \\ %
\approx {\bf 5+1}  \endSp $ & $\Sp 1+4  \\ +\left[ 1\right]  \\ ={\bf 5+1} 
\\ \approx {\bf 5}+{\bf 1}  \endSp $ & $\Sp 6  \\ +\left[ 4\right]  \\ ={\bf %
10}  \\ \approx {\bf 10}  \endSp $ & $\Sp \left( 4\right)  \\ _{move} 
\endSp $ & $\Sp \left( 1\right)  \\ _{move}  \endSp $ & $\frac{^{SL(5)}}{%
^{SO(5)}}$ \\ \hline
$_{\frac{5,1}{5,1}}$ & $\Sb \left( {\bf 4,4}^{*}\right)  \\ \left( {\bf 4}%
^{*},{\bf 4}\right)  \endSb $ & $_{\Sp 5+10  \\ +1  \\ ={\bf 1+15}  \\ %
\approx \left( {\bf 4,4}\right)  \endSp }$ & $_{\Sp  \\ 1+5+  \\ 5+\left[
1\right]  \\ =2\times {\bf 6}  \\ \approx \left( {\bf 0,5}\right)  \\ %
+\left( {\bf 5,0}\right)  \\ +2\left( {\bf 0,0}\right)  \endSp }$ & $_{\Sp %
1+10  \\ +\left[ 5\right]  \\ ={\bf 1+15}  \\ =\left( {\bf 4,4}\right) 
\endSp }$ & $_{\Sp 10^{+}  \\ +10^{-}  \\ ={\bf 10}^{+}  \\ +{\bf 10}^{-} 
\\ \approx ({\bf 10,}1{\bf )}  \\ +(1,{\bf 10})  \endSp }$ & $_{\Sp \left(
5\right)  \\ _{move}  \endSp }$ & $_{\Sp \left( 1\right)  \\ _{move}  \endSp %
}$ & $_{\frac{SO(5,5)}{\Sp SO(5)  \\ \otimes SO(5)  \endSp }}$ \\ \hline
$^{\frac{6,1}{4,1}}$ & $^{\left( {\bf 8,4}\right) }$ & $\Sp 6+15  \\ +6  \\ %
+\left[ 1\right]  \\ ={\bf 7+21}  \\ \approx {\bf 27}+{\bf 1}  \endSp $ & $
\Sp 1+6  \\ +15+\left[ 6\right]  \\ ={\bf 7+21}  \\ \approx {\bf 27+1} 
\endSp $ & $\Sp 1+20  \\ +\left[ 15\right]  \\ ={\bf 1+35}  \\ \approx {\bf %
36}  \endSp $ & $\Sp \left( 15\right)  \\ _{move}  \endSp $ & $\Sp \left(
6\right)  \\ _{move}  \endSp $ & $\Sp \left( 1\right)  \\ _{move}  \endSp $
& $^{\frac{^{E_{6(6)}}}{^{USp(8)}}}$ \\ \hline
$^{\frac{7,1}{3,1}}$ & $^{\Sp ({\bf 8}^{+}{\bf ,(2,0))}  \\ ({\bf 8}^{-}{\bf %
,}(0,{\bf 2)})  \endSp }$ & $\Sp 7+21  \\ +21  \\ +\left[ 7\right]  \\ {\bf %
=28+28}  \\ \approx {\bf 28}_c  \endSp $ & $\Sp 1+7  \\ +35  \\ +\left[
21\right]  \\ ={\bf 8+56}  \\ \approx {\bf 63+1}  \endSp $ & $\Sp 1^{\pm } 
\\ +35^{\pm }  \\ ={\bf 1}^{\pm }  \\ +{\bf 35}^{\pm }  \\ \approx {\bf 36}%
_c  \endSp $ & $\Sp \left( 21\right)  \\ _{move}  \endSp $ & $\Sp \left(
7\right)  \\ _{move}  \endSp $ & $^0$ & $\frac{^{E_{7(7)}}}{^{SU(8)}}$ \\ 
\hline
$^{\frac{8,1}{2,1}}$ & $^{\left( {\bf 16,2}\right) }$ & $\Sp 8+28  \\ +56 
\\ +\left[ 28\right]  \\ ={\bf 36+84}  \\ \approx {\bf 120}  \endSp $ & $\Sp %
1+8+70  \\ +\left[ 1+56\right]  \\ ={\bf 1+9}  \\ {\bf +126}  \\ \approx 
{\bf 135}  \\ {\bf +1}  \endSp $ & $\Sp \left( 1+56\right)  \\ _{move} 
\endSp $ & $\Sp \left( 28\right)  \\ _{move}  \endSp $ & $^0$ & $^0$ & $%
\frac{^{E_{8(8)}}}{^{SO(16)}}$ \\ \hline
\end{tabular}
\]

\[
\text{Table I. Classification of Q}_\alpha ^a\text{ and Z}_{\mu _1\cdots \mu
_p}^{ab}\text{ under 11D (or 12D) and }K\text{.} 
\]

What is the meaning of the $p$-form central extension $Z_{\mu _1\cdots \mu
_p}^{ab}$? Since this is a charge in a global algebra, there ought to exist
a ($p+1$)-form local current $J_{\mu _0\mu _1\cdots \mu _p}^{ab}\left(
x\right) $ whose integral over a space-like surface embedded in $d$%
-dimensions gives 
\begin{equation}
Z_{\mu _1\cdots \mu _p}^{ab}=\int d^{d-1}\Sigma ^{\mu _0}\,\,J_{\mu _0\mu
_1\cdots \mu _p}^{ab}\left( x\right) .
\end{equation}
The current couples to the fields of low energy physics (i.e. supergravity).
In the case of usual central charges that are Lorentz singlets $Z^{ab}$
(i.e. $p=0)\,$the current is associated with charged particles. Such a
current may be constructed as usual from worldlines (or equivalently from
local fields) as follows 
\begin{equation}
J_\mu ^{ab}\left( x\right) =\int d\tau \sum_iz_i^{ab}\,\delta ^d\left(
x-X^i\left( \tau \right) \right) \,\,\partial _\tau X_\mu ^i\left( \tau
\right) .
\end{equation}
The $z_i^{ab}$ are the charges of the particles labelled by $i.$ This
current couples in the action to a gauge field $A_{ab}^\mu $, and it appears
as the source in the equation of motion of the gauge field 
\begin{eqnarray}
\begin{array}{l}
S\sim \sum_i\int d\tau \,A_{ab}^\mu \left( X^i\left( \tau \right) \right)
\,\partial _\tau X_\mu ^i\left( \tau \right) \,\,z_i^{ab} \\ 
\,\,\,\,=\int d^dx\,\,A_{ab}^\mu \left( x\right) \,J_\mu ^{ab}\left( x\right)
\\ 
\partial _{\lambda \,\,}\partial ^{[\lambda }A_{ab}^{\mu ]}\left( x\right)
=J_{ab}^\mu \left( x\right) .
\end{array}
\end{eqnarray}

The generalization to the higher values of $p$ is straightforward$:$ In
order to have a charge that is a $p$-form we need a current $J_{\mu _0\mu
_1\cdots \mu _p}^{ab}\left( x\right) $ that is a $(p+1)$-form$.$ This in
turn requires a $p$-brane to construct the current, 
\begin{eqnarray}
J_{\mu _0\mu _1\cdots \mu _p}^{ab}\left( x\right) &=&\int d\tau d\sigma
_{1\cdots }d\sigma _p\sum_iz_i^{ab}\,\delta ^d\left( x-X^i(\tau ,\sigma
_{1,}\cdots \sigma _p)\right) \,\, \\
&&\times \partial _\tau X_{[\mu _0}^i\cdots \,\partial _{\sigma _p}X_{\mu
_p]}^i(\tau ,\sigma _{1,}\cdots \sigma _p)\,\,,  \nonumber
\end{eqnarray}
and its coupling to supergravity fields requires a $(p+1)$-form gauge
potential $A_{\mu _0\mu _1\cdots \mu _p}\left( x\right) $ such that 
\begin{eqnarray}
S &\sim &\int d^dx\,\,A_{ab}^{\mu _0\mu _1\cdots \mu _p}\left( x\right)
\,J_{\mu _0\mu _1\cdots \mu _p}^{ab}\left( x\right) \\
&=&\sum_i\int d\tau d\sigma _{1\cdots }d\sigma _p\,A_{ab}^{\mu _0\mu
_1\cdots \mu _p}\left( X^i\right) \,\,\partial _\tau X_{[\mu _0}^i\cdots
\,\partial _{\sigma _p}X_{\mu _p]}^i\,\,z_i^{ab},  \nonumber
\end{eqnarray}
and 
\begin{equation}
\partial _{\lambda \,\,}\partial ^{[\lambda }A_{ab}^{\mu _0\mu _1\cdots \mu
_p]}\left( x\right) =J_{ab}^{\mu _0\mu _1\cdots \mu _p}\left( x\right) .
\end{equation}
As is well known by now there are perturbative as well as non-perturbative
couplings of $p$-branes to supergravity in various dimensions. Hence the $%
Z_{\mu _1\cdots \mu _p}^{ab}$ are present in the superalgebra and they
correspond simply to the charges of $p$-branes. The classification of their $%
ab$ indices under duality groups is the subject of the next section, but
here we already see that there is a one to one correspondence between the $p$%
-forms $Z_{\mu _1\cdots \mu _p}^{ab}$ and the ($p+1)$-form gauge potentials $%
A_{ab}^{\mu _0\mu _1\cdots \mu _p}$ that appear as massless states in string
theory in the NS-NS or R-R sectors.

The main message is that from the point of view of the superalgebra all $p$%
-branes appear to be at an equal footing. Isometries of the superalgebra
that will be discussed below treats them equally and may mix them with each
other in various compactifications. The theory in $d$ dimensions has $(p+1)$%
-forms $A_{ab}^{\mu _0\mu _1\cdots \mu _p}$ which appear as massless vector
particles in the string version of the fundamental theory. These act as
gauge potentials and couple at low energies to charged $p$-branes. This
generates a non-trivial central extension $Z_{\mu _1\cdots \mu _p}^{ab}$ in
the superalgebra. The number of such central extensions ($ab$ indices) is in
one-to-one correspondence with the number of the $(p+1)$-forms $A_{ab}^{\mu
_0\mu _1\cdots \mu _p}$, and these numbers can be obtained by counting the
possible combination of (symmetric/antisymmetric) indices $ab$ associated
with the supercharges.

\section{Duality groups}

In the discussion above we concentrated on the 11D (or 12D) content of the
supercharges and the central extensions. We now turn to duality. In string
theory the T-duality group is directly related to the number of compactified
left/right string dimensions. Therefore, in our notation, for a string of
type II it is $T=SO(c,c).\,\,$Its maximal compact subgroup is $%
k=SO(c)_L\otimes SO(c)_R$ where $L,R$ denote left/right movers respectively.
The index $a$ on the supercharges $Q_\alpha ^a$ corresponds precisely to the
spinor index of $SO(c)_L\otimes SO(c)_R$ (see table III in \cite{ibyank}).
Investigating the supercharges listed in Table I shows that the index $a$
that was classified there under the hidden non-compact group $%
SO(c+1,1)_{hidden}$ can be reclassified under the perturbatively explicit
maximal compact subgroup $k\subset T$ of T-duality, $k=SO(c)_L\otimes
SO(c)_R.$ These two groups are not subgroups of each other, but they do have
a common subgroup $SO(c).$ Recall that $c$ is the number of compactified
dimensions (other than the two hidden dimensions), and $SO(c)$ is the
rotation group in these internal dimensions.

Next we look for the {\it compact} group $K$ that contains $SO(c)_L\times
SO(c)_R,\,$ $SO(c+1)$ and that has an{\it \ irreducible} representation for
the index $a$ (total dimension $N$)$.$ By virtue of containing $k\subset T$
the group $K\supset k$ must be related to a larger group of duality $U$ that
contains $T$. The groups $K$ and $U$ are listed in Table I. The subgroup
hierarchy that emerges is as follows

\[
\begin{array}{l}
\stackunder{\downarrow }{\stackunder{c\text{ }compact\,+\,2\text{ }hidden%
\text{ }dims.}{SO(c+1,1)}}\otimes \stackunder{spacetime}{SO(d-1,1)}%
\,\,\rightarrow \stackunder{\uparrow }{\left. 
\begin{array}{c}
a\text{ on }Q_\alpha ^a \\ 
ab\text{ on }Z_{\mu _1\cdots \mu _p}^{ab}
\end{array}
\right. } \\ 
\left. 
\begin{array}{l}
\left. 
\begin{array}{l}
SO(c+1)_{\,\,\,1\,hidden\,dim} \\ 
SO(c)_L\otimes SO(c)_R
\end{array}
\right\} \,\,\,\Longleftarrow \,\,\,\,\,\,\,\stackunder{(duality)}{\stackrel{%
maximal\,compact}{K}} \\ 
\,\,\,\,\,\,\,\,\,\,\,\,\,\,\,\,\,\,\stackrel{\uparrow }{\stackunder{%
(T-duality)}{SO(c,c)}}
\end{array}
\right\} \Longleftarrow \stackunder{(duality)}{U}
\end{array}
\]

Since the same $N$ dimensional basis of supercharges labelled by $a$ knows
about both duality and the hidden dimensions, this must provide a bridge for
relating properties of the states of the theory under both qualities. The
first consequence of this is the reclassification of the central extensions $%
Z_{\mu _1\cdots \mu _p}^{ab}.$ Previously they were related to 11D\ (or 12D)
as in Table I. But now the combination $ab$ corresponds to the symmetric or
antisymmetric product of the $N$ dimensional representation of $K.$
Therefore, {\it the central extensions are now also classified under} $\dot
K $. The result is the total dimension listed in Table I in bold numbers
following the $\approx $ sign. These numbers are indeed dimensions of
irreducible multiplets under $K.$

The main point is that the supercharges as well as the central extensions
are now classified under hidden (broken) symmetries of two different types.
The first one $SO(c+1,1)_{hidden}$ relates to 11 or perhaps 12 hidden
dimensions, and the second one $K\subset U$ relates to $U$-duality$.$ The
common compact subgroup $SO(c+1)$ already contains non-perturbative
information about the spacelike hidden dimension, but more information about
the hidden time-like dimension and about $U$-duality is contained in the
larger group structures.

\section{U-duality and non-perturbative states}

Under the assumption that the superalgebra is valid as a dynamical (broken)
symmetry in the entire theory, all states would belong to multiplets of the
(broken) superalgebra, including the central extensions and the $p$-branes
associated with them. One would then expect to be able to classify the
physical states of the theory according to different modules of $%
SO(c+1,1)_{hidden}$ and $K\subset U$ that have intersections with each other
in the form of (broken) $SO(c+1)$ multiplets. Each one of these
classifications contains non-perturbative states related to either duality
or hidden dimensions. By finding them and studying their couplings
consistent with the superalgebra one would be able to learn certain global
properties of the underlying theory.

The scheme for finding the non-perturbative states is as follows. First
identify the perturbative string states, classify them under supermultiplets
and identify their classification under the perturbatively explicit $%
SO(c)_L\otimes SO(c)_R.$ Then try to reclassify them under the bigger group $%
K.$ If additional states are needed to make complete $K$ multiplets add them
(these extra states are presumably $p$-branes, $D$-branes). There may be
non-unique ways of completing $K$ multiplets. If so, then try to make it
consistent with the presence of the hidden dimensions by making sure that
the $SO(c$+1) representations embedded in $K$ multiplets are consistent with
the structure of the central charges listed in the table. When this is
achieved one should also check that it is all consistent with a
compactification of a collection of states that starts in 11 dimensions,
i.e. consistency with 11-dimensional (broken) multiplets with signature
(10,1). One may need to add at this stage more non-perturbative states that
are not in the same $K$-multiplet with some perturbative string state
(presubambly more $p$-or $D$-brane states). So far one should expect
consistency with ``M-theory''. Finally, check if the structure of the
representations that emerge in this way is also consistent with 12
dimensions, with signature (10,2). In this way many properties of
non-perturbative states can be deduced.

Such a program was initiated in previous papers \cite{ib11d}\thinspace 
\cite{ibyank}\thinspace 
\cite{sendb}. The only central extension included in those discussions is
the $p=0$ case in various dimensions. Recall that the $p=0$ central
extension in lower dimensions contains pieces of the $p\geq 1$ central
extensions of higher dimensions. Therefore, the non-perturbative states
include p-branes in their internal dimensions. Their results, on the
consistency between (10,1) and U-duality is concerned, are summarized here.
More will be said elsewhere \cite{ibbeyondm}
about (10,2) and the other central extensions.

\section{U-duality and 11D}

\subsection{perturbative and non-perturbative states}

In the toroidaly compactified type II string on $R^{d-1,1}\otimes T^c,$ with 
$d+c=10,$ the perturbative vacuum state has Kaluza-Klein (KK) and winding
numbers, and is also labelled by the $2_B^7+2_F^7$ dimensional Clifford
vacuum of zero modes (in the Green-Schwarz formalism). The closed string
condition $L_0=\bar L_0$ can be satisfied without requiring equal excitation
levels $l_{L,R}$ for left/right movers. Hence the perturbative states are 
\begin{equation}
\begin{array}{l}
({\rm {Bose}\oplus {Fermi\ oscillators})_L^{\left( l_L\right) }} \\ 
\times ({\rm {Bose}\oplus {Fermi\ oscillators})_R^{\left( l_R\right) }} \\ 
\times \,\,|vac,\,\,p^\mu ;\vec m,\vec n>
\end{array}
\label{pertst1}
\end{equation}
where the $c$-dimensional vectors $\left( \vec m,\vec n\right) $ are the
Kaluza-Klein and winding numbers that label the ``{\it perturbative base}''.
These quantum numbers satisfy the relations 
\begin{eqnarray}
l_L+\frac 12\vec p_L^2 &=&l_R+\frac 12\vec p_R^2=M_d^2  \label{mass} \\
\vec p_R^2-\vec p_L^2 &=&\vec m\cdot \vec n=l_L-l_R  \nonumber
\end{eqnarray}
where $\vec p_{L,R}$ depend as usual \cite{t-review} on $\left( \vec m,\vec
n\right) $ and ($G_{ij},B_{ij})$ that parametrize the torus $T^c,$ while $%
M_d $ is the mass in $d$-dimensions $M_d^2=p_\mu ^2.$ By using the methods
of \cite{ib11d} we can identify the following supermultiplet structure for
the string states (\ref{pertst1}) at levels $\left( l_L,l_R\right) $ 
\begin{eqnarray}
\left( 0,0\right) &:&\left( 2_B^7+2_F^7\right) \otimes 1_L\otimes 1_R 
\nonumber \\
\left( 0,l_R\right) &:&\left( 2_B^{11}+2_F^{11}\right) \otimes 1_L\otimes
\sum_ir_{iR}^{(l_R)}  \nonumber \\
\left( l_L,0\right) &:&\left( 2_B^{11}+2_F^{11}\right) \otimes
\sum_ir_{iL}^{(l_L)}\otimes 1_R  \label{pertst} \\
\left( l_L,l_R\right) &:&\left( 2_B^{15}+2_F^{15}\right) \otimes
\sum_ir_{iL}^{(l_L)}\otimes \sum_ir_{iR}^{(l_R)}  \nonumber
\end{eqnarray}
The $2_B^{11}+2_F^{11}$ corresponds to the intermediate supermultiplet of
11D supersymmetry. The structures $\sum_ir_{iL,R}^{(l_{L,R})}$ are listed in
Table II up to level 5.
\begin{eqnarray*}
&& 
\begin{tabular}{|l|l|}
\hline
Level & SO(9)$_{L,R}$ reps $\left( \sum_ir_i^{(l_{L,R})}\right) _{L,R}$ \\ 
\hline
$l_{L,R}=1\quad $ & $1_B$ \\ \hline
$l_{L,R}=2$ & $9_B$ \\ \hline
$l_{L,R}=3$ & $44_B+16_F$ \\ \hline
$l_{L,R}=4$ & $(9+36+156)_B+128_F$ \\ \hline
$l_{L,R}=5$ & $
\begin{array}{l}
\left( 
\begin{array}{c}
1+36+44+84 \\ 
+231+450
\end{array}
\right) _B \\ 
+\left[ 16+128+576\right] _F
\end{array}
$ \\ \hline
\end{tabular}
\\
&&\text{Table II. L/R oscillator states in 10D.}
\end{eqnarray*}
The $SO(9)_{L,R}$ representations in this table are reduced to
representations of $SO(d-1)_{L,R}\otimes SO(c)_{L,R}.$ So, a general
perturbative string state is identified by ``index space'' and ``base
space'' in the form 
\begin{equation}
\phi _{indices}^{\left( l_Ll_R\right) }\left( base\right)  \label{states}
\end{equation}
The base space are the quantum numbers coming through the $\left( \vec
m,\vec n\right) $ and the indices are given by the product of
representations in (\ref{pertst}) and Table II 
(which may be extended beyond level 5).
These are all the
perturbative type II string states in $d$-dimensions.

The spectrum of the non-perturbative states is much richer. There are many
central charges in the supersymmetry algebra (see Table I) and those provide
sources that couple to the NS-NS as well as R-R gauge potentials. Therefore
one finds a bewildering variety of non-perturbative solutions of the low
energy field equations as examples of non-perturbative states that carry the
non-perturbative $p$-brane charges $Z_{\mu _1\cdots \mu _p}^{ab}$. We will
take an algebraic approach to describe them, by imposing the stucture of the
superalgebra discussed earlier. The base quantum numbers are now extended to
include the non-perturbative charges that appear in the global superalgebra
(here we concentrate on $p=$0-branes only, ignoring the higher $p$-branes in
this paper). 
\begin{equation}
|vac,\,\,p^\mu ;\vec m,\vec n,z^I>
\end{equation}
where the 0-brane charges are $\left( \vec m,\vec n,z^I\right) $. From the
point of view of string theory, the $z^I$ are non-perturbative charges that
couple to the R-R sector, while $\left( \vec m,\vec n\right) $ are the
perturbative charges that couple to the NS-NS sector. In the notation of
Table I we identify the generators that correspond to $\left( \vec m,\vec
n,z^I\right) $ as follows 
\begin{eqnarray*}
\vec m\,\, &\rightarrow &P^i,\,\,\quad \,\vec n\,\,\rightarrow
Z^{i,c+1},\quad \,\,z^I\,\,\rightarrow (P^{c+1},Z^{ij},X^{r_1\cdots r_5})\,
\\
i,j &=&1,2,\cdots ,c\quad \quad r_1=1,\cdots ,c,c+1
\end{eqnarray*}
That is, $\vec m$ corresponds to the Kaluza-Klein momenta excluding the
extra hidden coordinate, the winding numbers $\vec n$ correspond to the last
column or row of $Z^{r_1r_2},$ while all remaining 0-brane charges are
non-perturbative. Although there is a big asymmetry among these charges from
the point of view of the string, they are on equal footing from the point of
view of the superalgebra, and they are classified in higher multiplets of $%
SO(c+1,1)_{hidden}$ and of $K\subset U$ as discussed before. The multiplets $%
Z^{ab}=\left( \vec m,\vec n,z^I\right) $ form the {\it non-perturbative base 
}in $\phi _{indices}\left( base\right) $ \footnote{%
According to the dimensions of representations in Table I, the 0-brane $%
Z^{ab}=\left( \vec{m},\vec{n},z^I\right) $ seem to correspond to complete
linear representations of $U$ for all dimensions except for $d=3$ (when $%
U=E_{8(8)}).$ Similarly, higher $p$-branes $Z_{\mu _1\cdots \mu _p}^{ab}$ do
not generally form linear representations of $U$. Also they seem to form
complete representations of $SO(c,c)$ for all cases except for ($d=5,\,p=3)$%
,\thinspace ($d=3,4,$ $\,p=2)$. We interpret these observations to mean that
the base is not generally a {\it linear} representation of either $T$ or $U$
duality groups, but it is a {\it linear} representation of $K$ or $%
SO(c+1,1). $}.

There are two types of new non-perturbative states: those obtained by
applying string oscillators on the non-perturbative base and those that
cannot be obtained in this way, but which are required to be present to form
a basis for U-duality transformations. The second kind require the extension
of the indices such that complete $K$ multiplets are obtained. These are
needed as intermediate states in matrix elements of the superalgebra which
is assumed to be valid in the full theory. So, a general state in the theory
is identified at each $l_{L,R}$ as in (\ref{states}). Both the base and the
indices have non-perturbative extensions. The full set of states is required
to form a basis for U-duality transformations at each fixed value of $%
l_{L,R}.$ These states are not degenerate in mass, hence the idea of a
multiplet is analogous to the multiplets in a theory with broken symmetry.

The BPS saturated states are those with either $l_L=0$ or $l_R=0.$ Even for
BPS saturated states there are the two types of non-perturbative states.
Typically the non-perturbative indices occur for $l_L\geq 2,\,\,l_R=0$. For
the BPS saturated states one can derive an exact non-perturbative formula
for the mass by using the supersymmetry algebra with central charges. For
example for $(d=9,\ \  c=1)$ for a non-perturbative BPS state with
KK momentum $p_9=m/R$, winding $w=nR$ and non-perturbative eleventh
momentum $p_{c+1}=z/r$, we have 
$l_R=0,$ $l_L= mn,$ and a mass formula 
\begin{equation}
M={\frac 1{\sqrt 2}} |nR_{10}+\sqrt{m^2/R^2+z^2/r^2}|.
\end{equation}
where $m,n,z$ are quantized integers and $R,r$ are moduli.
The presence of the non-perturbative (quantized) $p_{c+1}$ is a new piece in
the mass formula that differs from the perturbative string BPS states. 
This formula is derived
from the superalgebra with the usual methods, but allowing for the
non-perturbative $p_{c+1}.$ A special case is the uncompactified theory
 in 10D, for
which the BPS states (called black holes in \cite{dualities}) have masses
proportional to the 11th momentum. Further generalizations involving other
non-trivial $z^I$ will be given elsewhere. For non BPS saturated states we
cannot give an exact mass formula.

What about the hidden dimensions? In the {\it uncompactified} theory
consider all the states, including their values of the non-perturbative 11th
momentum . In Fourrier space the fields $\phi _{indices}\left( x^\mu
,x^{11}\right) $ seem to be 11-dimensional. This is possible only if the
indices also have an 11D structure. At levels $l=0,1$ it has been known that
this is true for a long time for the usual string states, and this is
evident from Table II (level $l=1$ is a singlet times the factor 
$2_B^{15}+2_F^{15}$ which has 11D content).
At higher levels $l\geq 2$ the string states by
themselves do not have the 11D structure for the indices. The minimal
structure of indices that would be needed in an 11D theory was identified
for all levels. This minimal structure has a definite pattern for massive
states given by 
\begin{equation}
indices\Rightarrow \left( 2_B^{15}+2_F^{15}\right) \times R^{\left( l\right)
}.  \label{10dpert}
\end{equation}
The factor $2_B^{15}+2_F^{15}$ can be interpreted as the action of 32
supercharges on a set of $SO(10)$ representations $R^{\left( l\right) }$ at
oscillator level $l.$ For the minimal set of indices the factor $R^{\left(
l\right) }$ is of the form of a sum of SO(9) representations that make up
SO(10) representations. 
\begin{equation}
R^{\left( l\right) }=\sum_{l^{\prime }=1}^l\left( \sum_ir_i^{(l^{\prime
})}\right) _L\times \sum_{l^{\prime }=1}^l\left( \sum_ir_i^{(l^{\prime
})}\right) _R  \label{11dlorentz}
\end{equation}
Each term in the sum over $l^{\prime }$ looks like the string states in
Table II \cite{ib11d}. Only the highest term ($l^{\prime }=l)$ corresponds
to the perturbative string states of level $l.$ The remaining terms
correspond to non-perturbative states with quantum numbers isomorphic to
those listed in Table II at the given levels. The meaning of this pattern
has not been understood so far. Furthermore, in the complete theory there
may be more states beyond the minimal set displayed above.

What about 12D? Can the states discussed above also be classified under $%
SO(10,2).$ First, for $l=0$ the answer is yes, since the massless states
classified with the Poincare group are also a representation of the
conformal group. Then the 11D $2_B^7+2_F^7$ massless states classified
according to the Lorentz group SO(10,1) also form a basis consistent with
the conformal group SO(10,2). A more interesting case is the $l=1$ first
massive level states 2$_B^{15}+2_F^{15}.$ As mentioned above, at rest the
physical states come in complete $SO(10)$ multiplets, where $SO(10)$ is the
rotation group in 11-dimensions. From the point of view of $SO(10,2)$ we
would like to show that they come in complete multiplets of $%
SO(10,1)^{\prime }$ where the time like component is the hidden timelike
coordinate. Indeed this is true for the 2$_B^{15}+2_F^{15}$ states! This can
be explained as being a simple property of the first excited level and of
the supercharges, as follows: These states may be regarded as the simplest
massive supermultiplet created by applying all possible combinations of the
32 supercharges on a singlet vacuum. Since the vacuum is a singlet of $%
SO(10,1)^{\prime },$ and the supercharges form the spinor representation of $%
SO(10,1)^{\prime }$ (by virtue of being a representation of $SO(10,2)$),
then the classification of the states under $SO(10,1)^{\prime }$ follows
automatically from the products of the 32-dimensional spinor. This is an
interesting signal of the presence of a hidden timelike dimesion. In this
paper there will be no more discussion of higher excited levels from the
point of view of 12D

\subsection{Dualities and non-perturbative spectrum}

The perturbative string states involved in the T-duality transformations are
not all degenerate in mass. Therefore, T-duality must be regarded as the
analog of a spontaneously broken symmetry, and the string states must come
in complete multiplets despite the broken nature of the symmetry. It is well
known that $T=O(c,c;Z)$ acts linearly on the the $2c$ dimensional vector $%
\left( \vec m,\vec n\right) $. However it is important to realize that it
also acts on the indices of $\phi _{indices}$ in definite representations 
\cite{ibyank}. The action of $T$ on the indices is an induced $k$%
-transformation that depends not only on {\it all} the parameters in $T$ $\,$%
but also on the background $c\times c$ matrices ($G_{ij},B_{ij}$) that
define the tori $T^c$ . Since the states in the previous section are all in $%
k=O(c)_L\times O(c)_R$ multiplets, the $T$-duality transformations do not
mix perturbative states with non-perturbative states.

A U-multiplet contains both perturbative as well as non-perturbative
T-multiplets. Like the $T$-duality transformations, the U-duality
transformations act {\it separately} on the base and the indices of the
states described by (\ref{states}) {\it without mixing index and base spaces}%
. The action of $U$ on the base quantum numbers $(\vec{m},\vec{n},z_I)$ is a
linear transformation in a representation of same dimension as the
representation of $K$ listed in Table I \cite{ibyank}. The action on index
space is an{\it \ induced field-dependent gauge transformation in the
maximal compact subgroup }$K$, whose only free parameters are the global
parameters in $U.$ This $\left( U,K\right) \,$ structure extends the
situation with the $\left( T,k\right) $ structure of the T-duality
transformations described in the previous paragraph. The
logical/mathematical basis for this structure is induced representation
theory. The bottom line is that in order to have U-duality multiplets, in
addition to the non-perturbative base, {\it the ``indices'' on the fields in
(}\ref{states}{\it ) must form complete }$K${\it -multiplets}.

By knowing the structure of a U-multiplet we can therefore predict
algebraically the quantum numbers of the non-perturbative states by
extending the quantum numbers of the known perturbative states given in (\ref
{pertst}). The prediction of these non-perturbative quantum numbers is one
of the immediate outcomes of our approach.

\subsection{An example}

It is very easy to analyze the case $\left( d,c\right) =\left( 6,\,4\right) $
so we present it here as an illustration$.$ In this case the spin group is $%
SO(5)$ and there are $4$ internal dimensions$.$ The duality groups and index
spaces follow from Tables I,II and (\ref{pertst}). The relevant information
is summarized by 
\begin{equation}
\begin{array}{l}
U=SO(5,5),\quad K=SO\left( 5\right) \otimes SO\left( 5\right) \\ 
T=SO(4,4),\quad k=SO\left( 4\right) _L\otimes SO\left( 4\right) _R \\ 
l_{L,R}=1:\quad \left( \sum_ir_i^{(l_{L,R})}\right) _{L,R}=1_{L,R} \\ 
l_{L,R}=2:\quad \left( \sum_ir_i^{(l_{L,R})}\right) _{L,R}=9_{L,R} \\ 
\quad \quad \quad \quad \quad =5_{L,R}^{space}\oplus 4_{L,R}^{internal} \\ 
l_{L,R}=3:\quad etc.
\end{array}
\end{equation}
where the $9_{L,R}$ have been reclassified according to their space and
internal components. The reclassification is done also for the short ($%
2_B^7+2_F^7),$ intermediate ($2_B^{11}+2_F^{11})$ and long ($%
2_B^{15}+2_F^{15})\,$ supermultiplet factors. It is clear from this form
that the $k=SO\left( 4\right) _L\otimes SO\left( 4\right) _R$ structure
follows directly from the separate left/right internal components, while the
spin of the state is to be obtained by {\it combining} left and right
content of the space part.

Here I will discuss an example involving BPS states which is very similar to
another discussion on non-BPS states given in \cite{ibyank}. Let us consider
the BPS saturated states $\left( l_L\neq 0,\,\,l_R=0\right) .$ The base
quantum numbers in $\phi _{indices}^{\left( l_L,0\right) }\left( base\right) 
$ form the 16 dimensional spinor representation of $U=SO(5,5)$ 
\begin{equation}
base=\left( \vec{m},\vec{n},z^I\right) =16{\rm \ \ {\ of }\ \ SO(5,5)}
\label{the16}
\end{equation}
Among these the eight quantum numbers $\left( \vec{m},\vec{n}\right) $ are
perturbative, while the remaining eight $z^I$ are non-perturbative. 0-branes
that carry these quantum numbers provide the sources for the field equations
of the 8 massless NS-NS vectors and the 8 R-R vectors respectively. The
representation content of the indices in $\phi _{indices}^{\left(
l_L,0\right) }\left( base\right) $ is 
\begin{equation}
\begin{array}{l}
indices=(2_B^{11}+2_F^{11})\times \\ 
\quad \times \left[ 
\begin{array}{l}
\left( \sum_ir_i^{(l_L)}\right) _L \\ 
+non-perturbative
\end{array}
\right]
\end{array}
\end{equation}
where $(2_B^{11}+2_F^{11})$ is interpreted as the SUSY factor. The full set
of indices must form complete $K=SO\left( 5\right) _L\otimes SO\left(
5\right) _R$ multiplets for consistency with the general U-duality
transformation. It can be shown generally that the SUSY factor does satisfy
this requirement because the supercharges themselves are representations of $%
SO(5)_{spin}\times K$ \cite{ibyank}. Therefore, the remaining factor in
brackets must be required to be complete $SO(5)_{spin}\times K$ multiplets.

At level $l_L=1$ the piece $\sum_ir_i^{(1)}=1$ is a singlet, as seen in
Table II. Hence no additional non-perturbative indices are needed at this
level. At level $l_L=2$ the piece $\sum_ir_i^{(2)}=9_L=5_L^{space}\oplus
4_L^{internal}$ is classified under $SO(5)_{spin}\times SO\left( 4\right)
_L\otimes SO\left( 4\right) _R\,$ as 
\begin{equation}
\left( 5,\left( 0,0\right) \right) +\left( 0,\left( 4,0\right) \right) .
\end{equation}
Obviously, this is not a complete $SO(5)_{spin}\times SO\left( 5\right)
_L\otimes SO\left( 5\right) _R$ multiplet. Therefore, non-perturbative
indices must be added just in such a way as to extend the $\left( 4,0\right) 
$ of $k=SO\left( 4\right) _L\otimes SO\left( 4\right) _R$ into the $\left(
5,0\right) $ of $K=SO\left( 5\right) _L\otimes SO\left( 5\right) _R.$ That
is 
\begin{equation}
\left( 4_{int}\right) _L\rightarrow \left( 5_{int}\right) _L.  \label{4to5}
\end{equation}
This extension determines the required non-perturbative indices for this
case. Note that this amounts to extending the $9_L$ into a $10_L,$ and
similarly for right-movers 
\begin{equation}
9_{L,R}\rightarrow 10_{L,R}.  \label{9to10}
\end{equation}
This is precisely what was needed in section-1 in order to obtain
consistency with an underlying 11D theory \cite{ib11d}.

At all higher levels $l_{L,R}$ the requirement for complete $K-$multiplets
coincides precisely with the requirement of an underlying 11D theory.
Therefore the full set of indices are the same as those given in eq.(\ref
{11dlorentz}). The story is the same with the non-BPS-saturated states at
arbitrary $l_{L,R}.$ This result was found in \cite{ib11d} by assuming the
presence of hidden 11-dimensional structure in the non-perturbative type-IIA
superstring theory in 10D. In ref.\cite{ib11d} a justification for (\ref
{9to10}) could not be given. However, in \cite{ibyank} and in the present
analysis $U$-duality demands (\ref{4to5}) and therefore justifies (\ref
{9to10}), and similarly for all higher levels.

Therefore for this particular compactification on $R^6\otimes T^4,$
U-duality and 11D Lorentz representations imply each other.

A consistency check between U-duality and D-branes was reported in \cite
{sendb} and in this conference. It is of interest to compare that analysis
to ours. We find complete agreement at level $l_L=1.$ But at higher levels $%
l\geq 2$ our scheme requires more states than the D-brane degeneracy
computed in \cite{sendb}. In his case the states corresponding to the
non-perturbative indices were not considered, seemingly because the special
U-duality transformation he considered (interchanging the two 8's in the 16
of (\ref{the16})) has a trivial transformation on our index space (does not
go outside of the 4$_{int}^{L,R}$). We have seen that under more general
U-transformations the extra indices are needed both for U-duality multiplets
as well as for the 11D interpretation. Thus, the D-brane or other
interpretation of these extra states is currently unknown.

For $\left( d,c\right) =\left( 10,0\right) ,\left( 9,1\right) ,\left(
8,2\right) ,\left( 6,4\right) $ the analysis for $l_{L,R}=2,3,4,5$ produces
exactly the same conclusion as the 11D analysis. That is, $U$-duality
demands that the $SO(9)_L\otimes SO(9)_R$ multiplets $\sum_ir_i^{(l_{L,R})}$
should be completed to $SO(10)_L\otimes SO(10)_R$ multiplets. The minimal
completion (\ref{11dlorentz}) is sufficient in this case. Hence, in these
compactifications $U$-duality is consistent with a hidden 11D structure, and
in fact they imply each other.

On the other hand for the other values $(d,c)=\left( 7,3\right) ,\left(
5,5\right) ,\left( 4,6\right) ,\left( 3,7\right) $ the story is more
complicated. At various low levels we found that the minimal index structure
required to satisfy $U$-duality is different than the {\it minimal structure}
of 11-dimensional supersymmetry multiplets (\ref{11dlorentz}). If both
U-duality and 11D are true then there must exist an even larger set of
states such that they can be regrouped either as 11D multiplets or as
U-duality multiplets. Exposing one structure may hide the other one. In fact
we have shown how this works explicitly in an example in the case $\left(
7,3\right) $ at low levels $l_{L,R}$ \cite{ibyank}. However, it is quite
difficult to see if the required set of states can be found at all levels.

\section{Final remarks}

The basic assumption that we made is that the superalgebra is valid in the
sense of a (broken) dynamical symmetry for the full theory. By studying the
isometries of the superalgebra, including the central extensions, many of
the features of duality could be displayed while some new features became
apparent, including the following:

\begin{enumerate}
\item  The central extensions (and the supercharges) have a structure
consistent with two hidden spacetime dimensions, with an overall signature
(10,2).

\item  As a consequence of central extensions of the superalgebra, $p$%
-branes naturally become part of the fundamental theory, and their
interaction with $p+1$ forms in supergravity are deduced. These $p$-branes
contribute to the non-perturbative states demanded by $U$-duality and hidden
higher dimensions on an equal footing.

\item  The structure of $U$-duality in type II superstrings, the groups, the
non-perturbative states and their classifications emerge naturally from the
structure of the superalgebra. This is summarized by Table I. Furthermore,
one may start with perturbative string states, but then add non-perturbative
states that are needed in order to provide a basis for the underlying
superalgebra and its isometries. This is a method of finding at least some
of the a-priori unknown non-perturbative states.
\end{enumerate}

\end{document}